\documentclass{raa}
\usepackage{graphicx,times}
\usepackage{natbib}
\usepackage{amsmath}
\usepackage{float,color}
\usepackage{savefnmark,booktabs}
\usepackage{threeparttable,supertabular}

\begin{document}

\title{A Model of Fast Radio Bursts: Collisions Between Episodic Magnetic Blobs 
}

\volnopage{ {\bf XXXX} Vol.\ {\bf X} No. {\bf XX}, 000--000}
\setcounter{page}{1}

\author{Long-Biao Li\inst{1,2}, Yong-Feng Huang\inst{1,2}$^{*}$, Jin-Jun Geng\inst{1,2}, Bing Li\inst{1,2,3}  }

\institute{School of Astronomy and Space Science, Nanjing University, Nanjing 210046, China {\it hyf@nju.edu.cn}\\
    \and
          Key Laboratory of Modern Astronomy and Astrophysics (Nanjing University), Ministry of Education, Nanjing 210046, China \\
    \and
          Particle Astrophysics Division, Institute of High Energy Physics, Chinese Academy of Sciences, Beijing 100049, China \\
\vs \no
   {\small Received XXXX ; Accepted XXXX}
}

\abstract{
  Fast radio bursts (FRBs) are bright radio pulses from the sky with millisecond durations and Jansky-level flux densities.
  Their origins are still largely uncertain. Here we suggest a new model for FRBs. We argue that the collision of a white
  dwarf with a black hole can give birth to a transient accretion disk, from which powerful episodic magnetic blobs
  will be launched. The collision between two consecutive
  magnetic blobs can result in a catastrophic magnetic reconnection, which releases a large amount of free magnetic energy
  and forms a forward shock. The shock propagates through the cold magnetized plasma within the blob in the collision region,
  radiating through synchrotron maser mechanism, which is responsible for a non-repeating FRB signal.
  Our calculations show that the theoretical energetics, radiation frequency, duration timescale,
  and event rate can be well consistent with the observational characteristic of FRBs.
}

   \authorrunning{Li, Huang, Geng \& Li}            
   \titlerunning{A Model of FRBs: Collisions Between Episodic Blobs}  
   \maketitle

\section{Introduction}           
\label{sect:intro}

Fast radio bursts (FRBs) are intense transient flares with a high flux and millisecond duration at radio wavelengths.
The first FRB was discovered from a search of archival Parkes survey data \citep{Lorimer2007}.
Interestingly, there might be more than one class of FRBs: e.g. repeating and non-repeating ones \citep{Keane2016,Pala2017}.
To date, one repeating FRB and 29 non-repeating FRBs have been reported
\citep{Petroff2016}\footnote{A FRB catalog can be found at http://www.frbcat.org},
but their physical nature still remains unknown.
The non-repeating FRBs are found to be generally unresolved, whereas the repeating bursts of FRB 121102 are resolved with a temporal structure.
Most FRBs have high dispersion measures (DMs) of $300 \sim 1500\,\rm pc\,cm^{-3}$,
which are defined as the line-of-sight integral of the free electron number density.
These DMs typically exceed the contribution from the electrons in our Milky Way by a factor of $\sim 10$ \citep{Li2017}.
\citet{Lorimer2007} and \citet{Thornton2013} proposed that
the large DM should be largely attributed to the contribution from the ionized intergalactic medium (IGM),
and the DM contribution from the host galaxy is estimated as $\rm DM_{host}\leq 100\,pc\,cm^{-3}$,
which means that FRBs' redshifts would be in the range of $z \sim 0.3$~---~1.
Thus, FRBs seem to be of extragalactic or even cosmological origin \citep[e.g.][]{Caleb2016,Li2017}.
Fortunately, the extragalactic origin is confirmed by the repetition source FRB 121102,
which allows a precise sub-arcsecond localization and for the first time shows the association
with a host galaxy \citep{Chat2017,Marcote2017,Tend2017}.
The observed fluences and the cosmological redshift of $z = 0.193$ imply that
the FRBs from the source of 121102 have a typical energy of $E_{\rm iso} \sim 10^{39} \rm\, erg$ \citep{Chat2017},
if the bursts are isotropic.

FRBs' short durations ($\sim$ a few ms) and high brightness require that
their sources should be compact, and the emission should be coherent \citep{katz2014,Luan2014}.
There are a lot of progenitor models proposed to explain FRBs.
For non-repeating FRBs, the models include double compact star mergers \citep{Totani2013,Ming2015},
interaction of companions with the magnetic field of extragalactic pulsars \citep{Mottez2014},
collisions of asteroids with neutron stars (NSs) \citep{Geng2015},
collapses of supermassive NSs into black holes (BHs) \citep{Falcke2014,Zhang2014},
magnetar giant flares \citep{Kulkarni2014,Lyubarsky2014},
giant radio pulses from pulsars \citep{Connor2016,Cordes2016},
the inspiral of double NSs \citep{Wang2016},
and collisions between NS and white dwarf (WD) \citep{Liu2017}.
For repeating bursts, the proposed models include asteroids falling randomly onto NSs \citep{Dai2016,Bagchi2017},
intermittent accretion of materials by a NS from a WD companion \citep{Gu2016},
``cosmic comb'' model \citep{Zhang2017},
active remnant NS after binary NS mergers \citep{Yama2017},
episodic relativistic e$^\pm$--beam from an active galactic nucleus (AGN) interacting with a surrounding cloud \citep{Vieyro2017},
or active young remnants of magnetars \citep{Belo2017,Metzger2017}.
In our work, we focus on the the non-repeating FRBs.

We note that in BH X-ray binaries and AGNs, episodic jets have been observed frequently \citep[e.g.][]{Fender2004,Chat2009}.
Episodic jets are intermittent and in the form of discrete moving plasma blobs.
They could be generated through magnetohydrodynamical (MHD) processes as described by \citet{Yuan2009}.
In this paper, we argue that FRBs can be produced during the merger of a white dwarf with a intermediate-mass black hole.
The mass transfer from the WD to the BH can give birth to an transient accretion disk around the BH.
Due to shear and turbulent motion of the accretion flow, a flux rope system near the disk is expected.
When the equilibrium of the flux rope is broken due to the accumulation of energy and helicity,
episodic magnetic blobs are ejected. The collision between two blobs will lead to a catastrophic
magnetic reconnection, which then generate a non-repeating FRB via synchrotron maser emission.
Our article is organized as follows.
In Section 2, we describe the ejection process of episodic magnetic blobs briefly.
In Section 3, synchrotron maser emission in the plasma blobs is calculated, and the model results are
compared with observations.
Finally, our discussion and conclusions are presented in Section 4.

\begin{figure}
\centering
\includegraphics[width=0.45\textwidth]{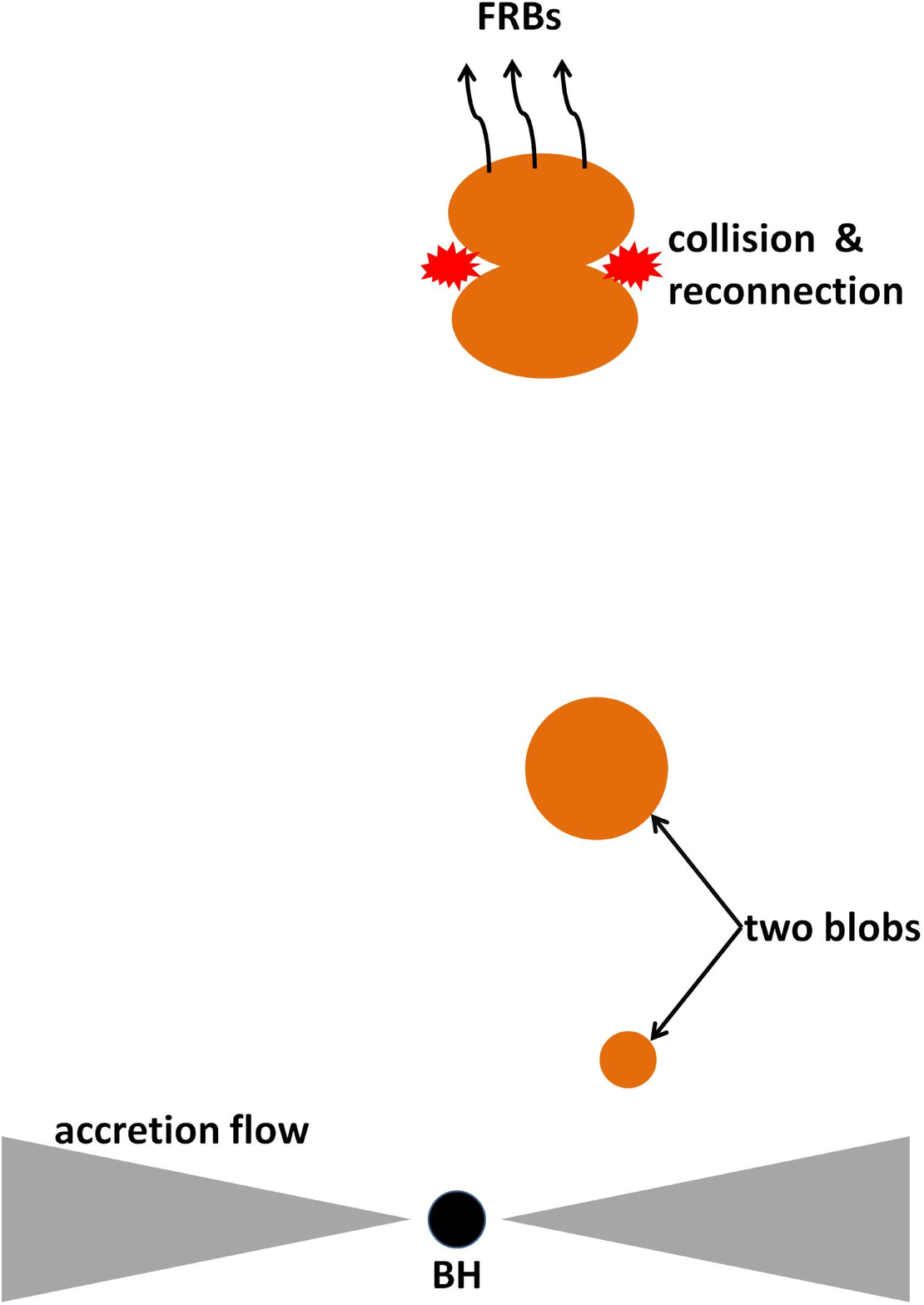}
\caption{Schematic illustration of the collision between two episodic blobs.
  Two consecutive magnetic blobs are ejected from a transient
  accretion disk. Their collision at a relatively large distance results in
  catastrophic magnetic reconnection.}
\label{fig1}
\end{figure}

\section{Ejection of Episodic Magnetic Blobs}

For a compact binary system consisting of a BH and a WD companion,
when the WD fills its Roche lobe, mass transfer can occur and material will
flow from the WD to the BH.
For a WD with enough a high mass, the mass transfer rate should be super-Eddington \citep{Dong2017},
which will trigger a runaway accretion process, leading the WD to merge with the BH.
In this case, a transient accretion disk can be formed around the BH.
It has been suggested that such a merging system can generate some kinds of
gamma-ray bursts \citep{Dong2017}. In our work frame, we argue that the transient
accretion disk can eject a few episodic magnetic blobs, which then give birth to
fast radio bursts via the collision between two adjacent blobs.

Accretion disks actually widely exist around black holes and other kinds of compact stars.
High speed wind from accretion disks can lead to the formation of a large scale corona around the accreting system.
According to \citet{Yuan2009}, closed magnetic field lines which emerge continuously from the accretion flow to the corona
are twisted and deformed due to the turbulence in the accretion flow, giving birth to a flux rope system in the corona.
With the accumulation of energy and tension, the equilibrium of the system would be broken when the threshold is reached.
The flux rope is thrust outward, generating an episodic jet.
As the accretion goes on, the above process repeats and a new blob will be produced.
In our modeling, two consecutive magnetic blobs moving relativistically at different speeds would
collide, resulting in the magnetic reconnection and leading to the release of a large amount of free
magnetic energy. The energy is dissipated via synchrotron maser to power the observed FRBs.
A schematic illustration of the overall picture of our model is shown in Fig. 1.

Let us consider a transient accretion disk surrounding a central BH with a mass of $M_{\rm{BH}} = 100 M_{\odot}$.
The disk is assumed to be an advection--dominated accretion flow \citep[ADAF, e.g.][]{Narayan1994,Abramo1995,Narayan1995,Chen2007}.
A flux rope system is expected to form in this accretion flow \citep{Yuan2009}.
Taking the mean mass accretion rate to be $\dot{M} = 10^{22}\rm\,g\,s^{-1}$, we then obtain
the temperature of the equatorial plane of the disk as \citep[e.g.][]{Narayan2001,Belo2003,Yuan2012}
\begin{equation}\label{eq1}
T_{c} = 9.2 \times 10^7 \alpha_{-1}^{-0.25} \dot{M}_{22}^{0.25} m_{2}^{-0.5} r^{-0.625} \rm\, K,
\end{equation}
where $\alpha_{-1}$ is the viscous parameter in units of 0.1, $m_2 = M_{\rm BH}/100 M_{\odot}$,
and $r = R/R_s$ is the dimensionless radius, with $R_{\rm s} = 2GM_{\rm BH}/c^2$ being the Schwarzschild radius.
Note that the convention of $Q = 10^x Q_{x}$ will be used throughout the paper.

Due to the topological structure of the magnetic field,
the available free magnetic energy is large in the flux rope region where the magnetic blob forms.
Similar to the coronal mass ejection in the Sun, the total available free magnetic energy of one blob
is $E_{\rm free} \approx 0.5 \times (1/12\, B_0^2\, V)$ \citep{Lin1998,Yuan2012}.
Here, $B_0$ is the magnetic induction intensity and $V = 4\pi R^3/3$ is the volume of the flux rope system.
\citet{Black2008} and \citet{Sorathia2012} have calculated the strength of the magnetic field with respect to
the thermal pressure of the gas ($P_{\rm gas}$) or the radiation pressure ($P_{\rm rad}$). They
defined a parameter $\beta = P_{\rm mag}/P$ to denote the ratio of the magnetic pressure over the gas or radiation pressure.
\citet{Sorathia2012} presented that a simple relation usually exists between $\alpha$ and $\beta$ as $\alpha/\beta \approx 0.5$.
For a radiation-pressure-dominated ADAF, $P_{\rm gas}$ is generally much smaller than $P_{\rm rad}$,
so that the gas pressure can be neglected and $\beta = P_{\rm mag}/P_{\rm rad}$.

For a given $\beta$, the magnetic induction intensity $B_0$ can be derived from
\begin{equation}
P_{\rm mag} = \frac{B_0^2}{8\pi} = \beta P_{\rm rad} = \beta \frac{4\sigma}{3c} T_c^4,
\end{equation}
which gives
\begin{equation}
B_0 = 7.2 \times 10^6 \alpha_{-1}^{-0.5} (\beta/0.2)^{0.5} \dot{M}_{22}^{0.5} m_2^{-1} (r/50)^{-1.25} \rm\, G.
\end{equation}
Thus, the available free magnetic energy of one blob near the accretion disk is
\begin{equation}
E_{\rm free} = 2.1 \times 10^{40} \alpha_{-1}^{-1} (\beta/0.2) \dot{M}_{22} m_{2} (r/50)^{0.5} \rm\, erg.
\end{equation}
We see that this energy is large enough to meet the requirement of the FRB energy budget.

When the flux rope system suddenly loses its equilibrium, a plasma blob can be ejected fiercely. The
blob is subsequently accelerated by the magnetic pressure gradient \citep{Tche2010,Kumar2015},
moving far away from the accretion flow.
Initially, the magnetic blob should be almost stationary and in the non-relativistic phase,
but its size can increase, i.e., it is expanding adiabatically with a speed of $\sim c$.
The non-relativistic phase lasts for the magnetic reconnection timescale at the base of the flux rope,
i.e., $t_{0} \sim 2 r_{\rm b}/v_{\rm rec} = 1.0 (v_{\rm rec}/10^{-2}c)^{-1} m_2 (r/50) \rm \, s$ \citep{Yuan2012},
where $r_b \sim  0.1 R \sim 3.0 \times 10^{8} m_2 (r/50)$ cm is the initial radius of the blob.
So, the size of the blob expands to $\Delta \sim c\,t_{\rm tec} \sim 3.0 \times 10^{10} m_2 (r/50)$ cm,
when the blob transits from the non-relativistic phase to the relativistic phase.

Fast reconnection leads to a decrease of the magnetic field with radius, which causes
effective acceleration of the plasma blob. At the distance of $\sim \Delta$, the
blob enters the relativistic phase with a typical Lorentz factor $\Gamma \sim \sigma_0^{1/3}$
\citep{Granot2011}, where $\sigma_0$ is the initial magnetization parameter at the base
of the accretion flow.

\section{Synchrotron Maser Emission From the Collision}

Let us consider two adjacent blobs ejected as mentioned above. We assume that the preceding one moves at a
smaller speed, while the latter has a higher bulk Lorentz factor. The faster blob will finally
catch up with the earlier slower one. Their collision will lead to a strong shock, just similar
to what happens in GRBs. These two blobs are initially separated
by $d = c\, \Delta t$ (with the faster one lagging behind the slower one),
where $\Delta t = z_{\rm rope}/v_A = 5.0 (\beta/0.2)^{-0.5} m_{2} (r/50)^{1.5} \,{\rm s}$ is the time interval
between the two consecutive blobs,
$z_{\rm rope}$ is the height of the flux rope
and is adopted as $z_{\rm rope} = 2.5\,R =3.7\times 10^{9}m_2(r/50)\rm\,cm$,
$v_A = B_0/\sqrt{4\pi\rho}$ is the Alfv\'{e}n speed,
and $\rho = 7.6 \times 10^{-6}\alpha_{-1}^{-1} m_{2}^{-2} \dot{M}_{22} (r/50)^{-1.5} \rm\,g\,cm^{-3}$
is the density of the corona/ADAF \citep[e.g.,][]{Horiuchi1988,Narayan1995,Yuan2012,Meyer2017}.
Assuming that the two blobs have different Lorentz factors of $\Gamma_{\rm fast}$ and $\Gamma_{\rm slow}$
($\Gamma_{\rm fast} > \Gamma_{\rm slow}$), the corresponding collision radius is
$r_{\rm col} \approx  2\, \Gamma^2\, c\, \Delta t$ \citep{Zhang2011}.
Note that hereafter, $\Gamma_{\rm fast}$ is shortened as $\Gamma$.
The collision of these two highly magnetized blobs are supersonic,
resulting in catastrophic magnetic reconnection in the area involved.
The reconnection releases a lage amount of magnetic energy and forms a shock wave,
which propagates through the magnetized, cold plasma within the blobs.

Due to the reconnection and turbulence, a large fraction of the magnetic energy is
converted into kinetic energies of particles,
which is responsible for particle acceleration and emission, finally powering the radio radiation.
The magnetic field in the comoving frame of the plasma blobs decreases as $B' \propto 1/z'$ \citep{Lyubarsky2009},
and the emission volume after the collision is $V_{\rm col} \approx f r_{\rm col}^2 (\Delta/\Gamma)$
(expressed in the observer's frame), where $f$ is the ratio of the solid angle of the emission region
with respect to $4 \pi$, and it can be adopted as $f \sim \frac{\Delta}{2 \pi r_{\rm col}}$.
For the emission occurring at a distance $r_{\rm col}$ from the central engine,
the energy release is comparable to $E_{\rm free}$.  
Together with equations 3) and 4), the Lorentz factor of the bubbles in the collision radius is
\begin{equation}
\Gamma \sim 42.1 (\beta/0.2)^{0.5} (r/50)^{-0.5}.
\end{equation}
The corresponding collision radius is thus
\begin{equation}
r_{\rm col} \approx 2\Gamma^2 c \Delta t= 5.3 \times 10^{14} (\beta/0.2)^{0.5} m_{2} (r/50)^{0.5} \rm\, cm.
\end{equation}

After the collision between the two consecutive blobs, the internal dissipation
process would form a forward shock that leads to synchrotron maser emission and shows up
as an FRB. As a viable radiation mechanism for FRBs, synchrotron maser emission has been extensively
discussed by many authors.
\citet{Lyubarsky2014} suggested that FRBs could result from the interactions of magnetic pulses
with the plasma within the nebula surrounding magentars. These interactions can produce
relativistic, magnetized shocks, leading to synchrotron maser emission and giving birth to FRBs.
\citet{Lu2017} also discussed possible conditions in which synchrotron maser emission can produce FRBs.
They suggested that the energy may come from the dissipation of free energy in an outflow,
which itself may be produced by the interaction between an external shock and the circum-stellar
medium in the forward shock region,
or by internal dissipation processes such as magnetic reconnections and collisions between shells.

In our modeling, we also consider the synchrotron maser emission as the main radiation mechanism.
Since the plasma within the blobs is highly magnetized, the forwardly shocked zone should also be
highly magnetized. Similar to \citet{Lyubarsky2014}, the inverse population is assumed to be
formed at the plasma energy levels of about $m_{e} c^{2} \Gamma$, and the synchrotron maser emission
predominantly proceeds at the Larmor rotation frequency of the plasma, i.e. $\nu'=eB'/(2\pi m_{e} c \Gamma)$.
Hence, the typical radiation frequency in the observer's frame can be estimated as
\begin{align}
\nu_{obs} & = \nu'\Gamma=\frac{eB'_{\rm col}}{2\pi m_{e} c}  \nonumber \\
          & = 1.2 \,\alpha^{-0.5} \dot{M}_{22}^{0.5} m_{2}^{-1} (r/50)^{-0.75}\rm\, GHz.
\end{align}

After the collision, the outflow is moving towards the observer with a Lorentz factor of $\Gamma$.
The scale of the maser emission area can be roughly estimated as $2\Delta/\Gamma^2$.
Thus the timescale of the maser emission, i.e., the observed FRB duration, is
\begin{equation}
t_{\rm FRB} \sim \frac{2\Delta}{\Gamma^2 c} \sim 1.2\, (\beta/0.2)^{-1} m_{2} (r/50)^{2} \rm\, ms.
\end{equation}

The DMs of FRBs are believed to be mainly contributed by the ionized IGM, and the
contribution from the local environment near the FRB engine should be small.
Let us estimate the intrinsic DM of the FRB source in our scenario.
Given that the central black hole mass is $\sim 100$~$M_{\odot}$ and the accretion rate is $\sim 10^{22}\rm\,g\,s^{-1}$,
the electron number density of the corona/ADAF can be roughly estimated as
$n_{e}= n_{p} = \rho/m_{p} \sim 4.6 \times 10^{18} \alpha_{-1}^{-1} m_2^{-2} \dot{M}_{22} (r/50)^{-1.5}\rm\, cm^{-3}$
\citep[e.g.,][]{Horiuchi1988,Narayan1995,Yuan2012,Meyer2017}.
According to the vertical density distribution of the corona, the electron density at the top of the flux rope is
$n_{e,\rm rope} \sim n_{e} {\rm exp}(-z_{\rm rope}^2/H_{\rm c}^2) = 1.3 \times 10^{11}\alpha_{-1}^{-1}
m_2^{-2} \dot{M}_{22} (r/50)^{-1.5} \,\rm cm^{-3}$,
in which $H_c$ refers to the scale height of the corona/ADAF and is roughly $0.6R$ \citep{Meyer2007,Kara2016,Qiao2017}.
Assuming that there is a wind outflow extending from the accretion system,
a conservative estimation on the DM contribution from the wind is
${\rm DM_{\rm Wind}} \sim \int_{z_{\rm rope}}^{\infty} n_{e,\rm rope} {\rm exp}(-z^2/z_{\rm rope}^2) dz
\approx 22.1 \alpha_{-1}^{-1} m_2^{-1} \dot{M}_{22} (r/50)^{-0.5}\rm\,pc\,cm^{-3}$.
On the other hand, the electron number density of blobs in the emission region is
$n_{e,\rm col}\approx n_{e,\rm rope} (2r_{b}/\Delta)^3=1.0 \times 10^{6} \alpha_{-1}^{-1}
m_2^{-2} \dot{M}_{22} (r/50)^{-1.5} \rm\, cm^{-3}$,
so that the DM contribution from the plasma within the blob itself can be calculated as
${\rm DM_{col}} \sim n_{e,\rm col} \Delta \approx 0.01 \alpha^{-1}_{-1} m_2^{-1} \dot{M}_{22} (r/50)^{-0.5}\rm\,pc\,cm^{-3}$.
Therefore, we see that the accretion disk system itself will only contribute a negligible
portion in the total dispersion measure in our model.

Recent studies suggest that the event rate of FRBs is in the range of 2000 -- 7000 $\rm Gpc^{-3}\, yr^{-1}$
within a maximum redshift of $z_{\rm max} = 1$ \citep{Li2017,Bhandari2018}, while the expected rate of WD-BH mergers
is $\sim 10^{4} \,\rm Gpc^{-3}\, yr^{-1}$ \citep{Cowper2015}.
Therefore, the event rate of FRBs is well consistent with that of WD-BH mergers.
From the above derivations, we see that the theoretical durations, typical radiation frequencies, energetics,
and event rate are all consistent with the observed features of FRBs.

\section{Discussion and Conclusions}

In this article, we argue that the non-repeating FRBs could originate from episodic magnetized plasma blobs ejected
from transient accretion disks around BHs. The transient disk can be formed
when a white dwarf merge with a 100$M_{\odot}$ black hole.
Due to the turbulence of the advection dominated accretion
flow around the intermediate-mass BH, some closed magnetic field lines will be twisted
and deformed continuously. They can emerge from the accretion disk and rise into the corona,
resulting in a huge flux rope system in the corona. When the magnetic energy of the flux rope
accumulates and reaches saturation, the system loses its equilibrium, ejecting a episodic magnetic blob,
whose free magnetic energy could ba as high as $2.1 \times 10^{40}$ erg.
In this way, a few episodic magnetic blobs can be launched from the transient accretion disk.
We show that the collision between two consecutive ejections can lead to catastrophic magnetic reconnection,
which releases a large amount of free magnetic energy and forms a forward shock.
The shock propagates through the magnetized, cold plasma within the blobs in the collision region,
which radiates via the synchrotron maser emission to produce a non-repeating FRB.
Our calculations suggest that the main observed features of FRBs, such as the
energetics, the radiation frequency, the duration, and the event rate, can all be satisfactorily explained.
Also, the intrinsic DM contribution from the accretion system itself is
negligibly small in our model.

In our modeling, episodic jets are generated by a transient ADAF disk.
The viscous timescale of the ADAF disk is $t_{\rm vis} = R/v_r = 10.4\, \alpha_{-1} m_2 (r/50)^{1.5} \rm\, s$,
where $v_r$ is the radial velocity. Comparing this expression with the timescale $\Delta t$ described
in Section 3, one can find $\Delta t < t_{\rm vis}$,
which means that in a typical viscous timescale, the flux rope system can eject 2 --- 3 magnetic bubbles.
We assumed that the blob is initially at rest, and the non-relativistic timescale is $t_0 \sim 2\, r_{\rm b}/v_{\rm rec}$.
However, both the initial size of the bubble and the reconnection velocity are poorly constrained,
so $\Delta$ can only be roughly estimated.
Meanwhile, the time interval between two adjacent blobs increases
with the increase of $r$ in the accretion disk, which is $\Delta t \propto r^{1.5}$.
It means that for a smaller $r\leq 10$, the condition of $\Delta t \leq t_0$ would be met,
which leads the two blobs to collide in the non-relativistic phase. 
On the other hand, a larger $r$ will allow for a longer $\Delta t$ and a larger $\Delta$,
which may generate a much larger $r_{\rm col}$ and a correspondingly lower magnetic strength at the collision radius.
This would lead the radiation frequency to be significantly lower than the observed
frequencies of typical FRBs.

It is interesting to note that \citet{Scholz2017} recently tried to search for the persistent X-ray
counterpart of FRB 121102. They finally reported an upper limit on the persistent X-ray luminosity at the level of
$3 \times 10^{41}\,\rm erg\,s^{-1}$ \citep{Scholz2017}. We thus need to examine whether there is any
persistent X-ray emission above this upper limit in our modeling.
Let us consider the release of the potential energy of the accreted material in our scenario.
The power can be easily estimated as $L \simeq \frac{GM\dot{M}}{3R_{s}} = 1.5 \times 10^{42}\rm\,erg\,s^{-1}$
by taking $M = 100$~$M_{\odot}$ and $\dot{M} = 10^{22}\rm\,g\,s^{-1}$,
where the innermost stable circular orbit is assumed to be at $3R_{s}$.
This value seems to be higher than the constraint presented by \citet{Scholz2017}.
But if we assume a reasonable efficiency of $\leq 0.1$ for converting the potential energy into
X-ray emission \citep{gruzinov1998}, then the X-ray luminosity will be less than $1.5 \times 10^{41}\rm\,erg\,s^{-1}$ and will
not conflict with the observational limit. Most importantly, note that the
ADAF disk in our scenario is a transient disk.
Once the accretion process stops and the accretion disk disappears, no X-ray emission
will be generated at all. Thus we basically do not expect any persistent X-ray emission
in our model. It is consistent with the observational constraint by \citet{Scholz2017}.

We have considered the ADAF disk as the source for episodic jets in our work.
However, it is still possible that the high accretion rate system is
the neutrino-dominated accretion flow (NDAF).
For an NDAF, the time interval between two blobs is usually longer than that in ADAF,
so that the flux rope system may eject only one blob in an stable timescale, i.e.,
$t_{\rm vis,NDAF}/\Delta t_{\rm NDAF} \propto 0.8\, (r/50)^{-0.35}$.
In this case, no FRBs could be generated.

Although episodic jets have been observed in many BH systems such as X-ray binaries and AGNs,
it is essentially difficult to image them due to their small sizes. With the large diameter radio
telescopes available, such as the Five-hundred-meter Aperture Spherical radio Telescope (FAST) \citep{Nan2011},
it is expected that the FRB sample can increase at a rate of $\sim 5$ everts per 1000 hours of observation time \citep{Li2017}.
When more FRBs are observed and localized, we may be able to get more useful information on these interesting
central engines that launch episodic magnetic blobs.

\normalem
\begin{acknowledgements}

We thank the anonymous referee for helpful suggestions that lead to a significant improvement
of this study.
This work is jointly supported by
the National Natural Science Foundation of China (Grant No. 11473012),
the National Basic Research Program of China (``973'' Program, Grant No. 2014CB845800),
the National Postdoctoral Program for Innovative Talents (Grant No. BX201700115),
the China Postdoctoral Science Foundation funded project  (Grant No. 2017M620199),
and by the Strategic Priority Research Program of the Chinese Academy of Sciences
``Multi-waveband Gravitational Wave Universe'' (Grant No. XDB23040000).
\end{acknowledgements}

\label{lastpage}
\end{document}